\documentstyle[preprint,aps]{revtex}
\begin{document}
\draft
\preprint{\vbox{
\hbox{IFT-P.42/97}
\hbox{hep-ph/yymmddd}
\hbox{July 1997}
}}
\title{
The electromagnetic gauge invariance in $SU(3)_L\otimes U(1)_N$ 
models of electroweak unification reexamined} 
\author{V. Pleitez$^{(a)}$ and M. D. Tonasse$^{(b)}$ }  
\address{
$^{(a)}$ Instituto de F\'\i sica Te\'orica\\
Universidade Estadual Paulista\\
Rua Pamplona 145\\ 
01405-900 -- S\~ao Paulo, SP\\Brazil} 
\address{$^{(b)}$ Instituto de F\'\i sica da Universidade do Estado 
do Rio de Janeiro\\
Rua S\~ao Francisco Xavier 524\\
20550-013 -- Rio de Janeiro, RJ\\
Brazil}
\date{\today}
\maketitle
\newpage
\begin{abstract}
Two models with 
$SU(3)_C\otimes SU(3)_L\otimes U(1)_N$ gauge symmetry 
are considered. 
We show that the masslessness of the photon does not prevent the
neutrinos from acquiring Majorana masses. That is, there is no relation 
between the VEVs of Higgs fields and the electromagnetic gauge invariance
contrary to what has been claimed recently. 

\end{abstract}
In model building of the electroweak interactions among elementary 
particles it is usually assumed that any neutral scalar field can 
get a Vacuum Expectation Value (VEV).
This assures the conservation of the electric
charge and the Abelian gauge invariance needed for maintaining the 
photon massless. 

In general the pattern of the symmetry breaking is not arbitrary but
depends on the structure of the theory, in particular on the
group representation content of the scalar fields~\cite{lfli}. For 
instance consider the case of $SU(2)$ gauge theory with a complex
doublet of scalar fields $\varphi$. In the unitary gauge the
vacuum expectation value (VEV) is $\langle\varphi\rangle=(0,\; v)^T$. 
Then, the original $SU(2)$ symmetry is completely broken. On the
other hand, if we have instead a real triplet $\phi$ with 
$\langle\phi\rangle=(0,\;0,\;v)^T$ (and we can always chose this vacuum
state), the gauge symmetry $SU(2)$ will be broken down to a residual
$U(1)$ gauge symmetry with one massless vector boson remaining.
In general to break the $SU(n)$ symmetry completely we need $n-1$ sets
of vector representations~\cite{lfli}. Charged scalars will spoil the 
electromagnetic gauge invariance in a trivial way. Hence, all discussions
go around neutral scalar bosons which can, in principle, gain a VEV. 
It is not obvious that with the proliferation of Higgs fields at
least one massless gauge boson, which can be identified as a photon,
remains. So, the general recipe for model building~\cite{bl} has to be
checked case by case. For instance, in the case of $SU(2)\otimes U(1)$ theory
one doublet with vacuum state $(0,\;v)^T$ chooses the $W^3$ direction so, if
we add a triplet with vacuum state $(0,\,v,\,0)^T$ the photon will gain a mass.
However, this is only an apparent effect since it is always possible to choose 
the vacuum state of the triplet $(0,\,\,0,\,v)^T$ in such a way that both vacuum 
states, the doublet's and the triplet's ones, are always in the same 
direction. 
Although this is always possible with a minimal set
of scalar multiplets, there are models in which the scalars needed
for breaking down the symmetry are not enough to generate all fermion masses.
New scalar multiplets are then necessary to generate the fermionic masses
and to have still the appropriate symmetry breaking.
Hence, we can wonder if in this case that feature will continue to be valid.
In fact, it has been claimed by \"Ozer~\cite{ozer} and Pisano {\it et al.}
~\cite{pupi}, that in some models based on 
$SU(3)_C\otimes SU(3)_L\otimes U(1)_N$ symmetry (3-3-1 models)
some neutral components of one sextet of scalar fields cannot gain a VEV since 
in this case it will break the electromagnetic $U(1)$ gauge symmetry 
generating a massive photon. The goal of this paper is to show that this 
is not true in any of those models.  

If we consider only leptons with electric charge $0$ and $\pm1$, the 
extensions of the standard model based on  the 
$SU(3)_C\otimes SU(3)_L\otimes U(1)_N$ gauge symmetry can be divided into two 
classes accordingly if there are two charged~\cite{pp,fhpp} or two neutral 
leptons~\cite{mpp} in each leptonic multiplet. In each class there are several
versions according to the number of singlets that are added. 
We consider a typical model
of the first class (Model A) and a model of the second class (Model B). 

The covariant derivatives of the triplet $\phi$ are
\begin{mathletters}
\label{dc}
\begin{equation}
{\cal D}_\mu\phi_i  =  \partial_\mu\phi_i - ig\left(\vec W_\mu 
.\frac{\vec\lambda}{2}\right)^j_i\phi_j - 
ig^\prime N_{\phi_i}B_\mu\phi_i,
\label{dct}
\end{equation}
and for the symmetric tensor $S$ we have
\begin{equation}
{\cal D}_\mu S_{mn}  = \partial_\mu S_{mn} - ig\left[\left(\vec W_\mu 
.\frac{\vec\lambda}{2}\right)^k_mS_{kn} + \left(\vec W_\mu .
\frac{\vec\lambda}{2}\right)^k_nS_{km}\right] - ig^\prime NB_\mu S_{mn};
\label{dcs}
\end{equation}
\end{mathletters}

In general, there are several distinct models depending on the way the electric
charge operator is embedded in the ge\-ne\-ra\-tors of 
$SU(3)_L\otimes U(1)_N$. We will use the leptonic sector, with the 
usual charges, $0$ and $\pm1$, to define the representation content of 
several 3-3-1 models. 

Model A is defined by the charge operator
\begin{mathletters}
\label{models}
\begin{equation}
Q_{{\rm A}}=\frac{1}{2}\left(\lambda_3-\sqrt3\lambda_8 \right)+N,
\label{q1}
\end{equation}
In this sort of models, 
besides the quarks of charge 2/3 and $-1/3$, there are exotic quarks 
with charges 5/3 and $-4/3$.

On the other hand, Model B is characterized by the charge operators 

\begin{equation}
Q_{{\rm B}}=\frac{1}{2}\left(\bar\lambda_3+\frac{1}{\sqrt3}
\bar\lambda_8 \right)+N,
\label{q2}
\end{equation}
\end{mathletters}
with $\bar\lambda=-\lambda^*$. In this kind of models
all quarks have charges 2/3 or $-1/3$.

In both types of models the Spontaneous Symmetry Breakdown (SSB)
\begin{equation}
SU(3)_L\otimes U(1)_N\longrightarrow SU_L(2)
\otimes U(1)_Y\longrightarrow U(1)_{\rm Q},
\label{ssb}
\end{equation}
is obtained by at least two scalar triplets. However, in order to give
mass to the quarks a third scalar triplet must be added or, in some 
cases, for giving a mass to the leptons a scalar sextet is also required. 
These multiplets contribute to the masses of the charged and neutral 
vector bosons too.

We have chosen the $\lambda$'s matrices in such a way that the first two 
entries in each triplet or antitriplet are the $SU(2)_L$ particles while 
the third one may be a new field. 

In Model A, Eq.~(\ref{q1}) implies
\begin{equation}
\Psi_{({\rm A})lL}=\left(
\begin{array}{c}
\nu_l \\ l^- \\ l^+
\end{array}\right)_L\sim({\bf3},0),\quad l=e,\mu,\tau.
\label{l1}
\end{equation}
Notice that the lepton multiplet in Eq.~(\ref{l1}) is that of 
Ref.~\cite{pp}. Since we are concerned here with the SSB aspect we will write
explicitly only the scalar multiplets.  
In this model, the introduction of right-handed 
neutrinos is optional.

Leptons with electric charge $0$ and $\pm1$ in model B are according
to Eq.~(\ref{q2})

\begin{equation}
\Psi_{({\rm B})lL}=\left(
\begin{array}{c}
\nu_l \\ l^- \\ \nu^c_l
\end{array}\right)_L\sim ({\bf3},-1/3).
\label{l2}
\end{equation}
In this sort of models, the right-handed charged leptons are in singlets
of the gauge symmetry. Neutrinos can, in principle, get 
a Dirac or Majorana mass, depending on the representation content of the 
scalar sector and also, on the VEV of the neutral scalar fields implemented 
by an appropriated scalar potential.

In all these models one of the quark families transforms in the same 
way as leptons. It is not necessarily the first family which does so. The 
distinction among families will be necessary only after the 
diagonalization of the quark mass matrices. Since we are concerned here only
with weak eigenstates, we do not need to distinguish which family transforms 
in the same way than the leptons.
This will be manifested only in the parametrization of the mixing matrices
$V^U_L$, $V^D_L$ and  $V^U_R,V^D_R$.
However, we stress that from the phenomenological point of view it will be 
important to consider if it is the lightest family, or the heaviest one, which 
has its components formed mainly by those fields transforming in the same way 
than the leptons.

For model A the scalar triplets are 
\begin{equation}
\eta=\left(
\begin{array}{c}
\eta^0 \\ \eta^-_1 \\ \eta^+_2
\end{array}
\right)\sim({\bf3},0),\quad
\rho=\left(
\begin{array}{c}
\rho^{+}\\ \rho^0 \\ \rho^{++}
\end{array}
\right)\sim({\bf3},+1),\quad
\chi=\left(
\begin{array}{c}
\chi^-\\ \chi^{--} \\ \chi^0
\end{array}
\right)\sim({\bf3},-1).
\label{t1}
\end{equation}
and the sextet from Eq.~(\ref{l1}) is
\begin{equation}
S=\left(
\begin{array}{ccc}
S^{0}_{11}\; & S^-_{12}/\sqrt2\; & S^+_{13}/\sqrt2\; \\
S^-_{12}/\sqrt2\; &  S^{--}_{22}\; & S^0_{23}/\sqrt2\; \\
S^+_{13}/\sqrt2 \;& S^0_{23}/\sqrt2\; & S^{++}_{33}
\end{array}
\right)\sim({\bf6},0).
\label{s1}
\end{equation}
We will use the notation  $\langle \eta^0\rangle\equiv v_\eta$,  
$\langle \rho^0\rangle\equiv v_\rho$,  $\langle \chi^0\rangle\equiv v_\chi$, 
$\langle S^0_{11}\rangle\equiv v_1$, $
\langle S^0_{23}\rangle\equiv \sqrt{2}v_2$.

From Eqs.~(\ref{t1}) and (\ref{s1}) we obtain (up to a $g^2/4$ factor)
\begin{equation}
M^2_{\rm A}=\left(
\begin{array}{ccc}
 v_\eta^2+v_\rho^2 +4v_1^2+2v^2_2& 
\frac{1}{\sqrt3}\left( v_\eta^2-v^2_\rho+4v_1^2+2v_2^2\right) &
 -2tv_\rho^2 \\
 \frac{1}{\sqrt3}\left(v_\eta^2-v^2_\rho+4v_1^2+2v^2_2\right) & 
\frac{1}{3}\left(v_\eta^2+v_\rho^2+4v_\chi^2+4v_1^2+2v^2_2\right) &
  \frac{2t}{\sqrt3}\left(v_\rho^2+2v_\chi^2\right)\\
-2tv_\rho^2 & \frac{2t}{\sqrt3}\left(v_\rho^2+2v_\chi^2\right) & 
4t^2\left(v_\rho^2+v_\chi^2\right)
\end{array}
\right).
\label{ma1}
\end{equation}

For model B the scalars are $\eta$ and 
$\sigma$:

\begin{equation}
\eta=\left(
\begin{array}{c}
\eta^0 \\ \eta^-_1 \\ \eta^-_2
\end{array}
\right)\sim({\bf3},-2/3),\quad
\sigma=\left(
\begin{array}{c}
\sigma^+ \\ \sigma^0_1 \\ \sigma^0_2
\end{array}
\right)\sim({\bf3},+1/3).
\label{t2}
\end{equation}

In the present model the sextet has the following charge attribution
\begin{equation}
S=\left(
\begin{array}{ccc}
S^{++}_{11}\; & S^+_{12}/\sqrt2\; & S^+_{13}/\sqrt2\; \\
S^+_{12}/\sqrt2\; &  S^{0}_{22}\; & S^0_{23}/\sqrt2\; \\
S^+_{13}/\sqrt2\; & S^0_{23}/\sqrt2\; & S^{0}_{33}
\end{array}
\right)\sim({\bf6},+2/3),
\label{s2}
\end{equation}
with $\langle S^0_{22}\rangle\equiv v_1$, 
$\langle S^0_{33}\rangle\equiv v_2$,
$\langle S^0_{23}\rangle\equiv \sqrt{2}v_3$.
We will use the notation 
$\langle \eta^0\rangle\equiv v_\eta$, $\langle \sigma_1^0\rangle\equiv 
v_{s1}$ and $\langle \sigma_2^0\rangle\equiv v_{s2}$.
 
From the Higgs multiplets in Eqs.~(\ref{t2}) and (\ref{s2}) we obtain the mass 
matrix (up to a $g^2/4$ factor)
\begin{mathletters}
\label{ma3}
\begin{equation}
M^2_B=\left(
\begin{array}{ccc}
m_{11} & \frac{1}{\sqrt3}m_{12} & -\frac{8t}{\sqrt3}m_{13} \\
\frac{1}{\sqrt3}m_{12} & \frac{1}{3}m_{22} & \frac{2t}{3\sqrt3}m_{23} \\
-\frac{2t}{\sqrt3}m_{13} & \frac{2t}{3\sqrt3}m_{23} & \frac{4t^2}{9}m_{33}
\end{array}\right),
\label{ma2}
\end{equation}
where
\begin{equation}
m_{11} =  v_\eta^2 + v_{s1}^2 + 4v_1^2 + 2v_2^2,\;\;
m_{12}  =  -v_\eta^2 + v_{s1}^2 + 4v_1^2 - 2 v_2^2,
\label{ma32}
\end{equation}
\begin{equation}
m_{13}  =  v_\eta^2/2 + v_{s1}^2/4 + v_1^2 + 2v_2^2,\;\;
m_{22}  =  v_\eta^2 + v_{s1}^2 + 4v_{s2}^2 + 4v_1^2 + 2v_2^2 + 16v_3^2,
\label{ma33}
\end{equation}
\begin{equation}
m_{23}  =  2v_\eta^2 - v_{s1}^2 + 2v_{s2}^2 - 4v_1^2 + 4v_2^2 + 8v_3^2,\;\;
m_{33}  =  4v_\eta^2 + v_{s1}^2 + v_{s2}^2 + 4v_1^2 + 8v_2^2 + 4v_3^2.
\label{ma2def}
\end{equation}
\end{mathletters}

It is straightforward to verify that both matrices in Eqs.~(\ref{ma1}) and
(\ref{ma2}) are singular {\it i.e.}, $\det M^2=0$ even if all VEVs are different from 
zero. Hence, contrary to what has been claimed 
in Refs.~\cite{ozer} and \cite{pupi} we have the photon massless as usual even
if all the neutral components of the sextets in Eqs.~(\ref{s1}) and
(\ref{s2}) gain a VEV.

In model B without the sextet, neutrino masses come only through 
Yukawa couplings to the triplet $\eta$:  $f_{lm}\epsilon^{ijk}
\overline{\Psi^c_{liL}}\Psi_{mjL}\eta_k$ being $f_{lm}$ antisymmetric matrix
in the flavor space. So, the neutrino mass 
spectrum is: one is  massless and the other two are mass degenerate.
However, since neutrinos and antineutrinos are placed in this model in the 
same multiplet, the mass degeneracy can be broken by radiative corrections, 
inducing Majorana masses if we choose an appropriate scalar 
potential~\cite{barbieri}. Or, we can make 
$\nu^c_L\to {\cal N}^c_L$ in Eq.~(\ref{l2}), being ${\cal N}$'s 
heavy neutral leptons and also adding the respective right-handed components 
$\nu_R$'s and ${\cal N}^c_R$'s. In this situation all 
neutral leptons gain an arbitrary Dirac mass via the Yukawa interaction with 
the triplets. This also happens in model A which include heavy charged 
leptons~\cite{pt}. 

Summarizing, in both kind of models it is possible all the VEVs of the sextet
be non-zero giving mass to the charged or neutral leptons and there is not a 
relation between any of the VEVs of 
those sextets and the breaking of the electromagnetic gauge invariance as
claimed in both Refs.~\cite{ozer} and \cite{pupi} and some time before
in Ref.~\cite{mpp}. We have verified also that this does not depend either on 
the specific model. For instance the model of Ref.~\cite{phf} is equivalent
to model A and the same results are obtained: the charge operators can be
related by a unitary 
transformation. In fact, by using the 
$SU(3)$ matrix
\begin{equation}
U=\left(
\begin{array}{ccc}
0 & 1 & 0 \\
-1 & 0 & 0 \\
0 & 0 & 1
\end{array}
\right),
\label{matrix}
\end{equation}
the operator $Q_A$ in Eq.~(\ref{q1}) turns into the charge operator defined in
Ref.~\cite{phf}. Similar equivalence occurs in models of the type of 
Model B that have been treated in literature as in Ref.~\cite{liu}.
 

\acknowledgements

We thank (V.P.)  Con\-se\-lho Na\-cio\-nal de 
De\-sen\-vol\-vi\-men\-to Cien\-t\'\i \-fi\-co e 
Tec\-no\-l\'o\-gi\-co (CNPq) for partial financial support and IF-UERJ and
IFT-UNESP (M.D.T.) for hospitality.

\end{document}